# Observation of gapless corner modes in synthetic translation dimensions

Wen-Jin Zhang[†], Hao-Chang Mo[†], Wen-Jie Chen, Xiao-Dong Chen[*], Jian-Wen Dong[*]

*School of Physics & State Key Laboratory of Optoelectronic Materials and Technologies,*
*Sun Yat-sen University, Guangzhou 510275, China.*

[†]These authors contributed equally to this work

[*]Corresponding author: chenxd67@mail.sysu.edu.cn, dongjwen@mail.sysu.edu.cn

## ABSTRACT

The introduction of synthetic dimensions in topological photonic systems enriches the exploration of topological phase of light in higher-dimensional space beyond three-dimensional real-space. Recently, the gapless corner modes of topological photonic crystals under translational deformation have been proposed, but their experimental observation is still absent. Here, we observe the gapless corner modes in a photonic crystal slab under translational deformation. The corner mode exhibits a frequency dependence that can be tuned through the translation of the slab. Importantly, we find that the existence of gapless corner modes is independent of the specific corner configuration. The gapless corner modes are experimentally imaged via the near-field scanning measurement, and validated numerically by full-wave simulations. Our work contributes to the advancement of topological photonics and provides valuable insights into the exploration of gapless corner modes in synthetic dimensions.

**Introduction**

Topological photonic crystals (TPCs) have emerged as a rapidly growing field of research in the past few years, holding great potential for revolutionizing the way we mold the flow of light [1-8]. A key characteristic of TPCs is the presence of topologically protected boundary modes localized at the crystal's boundary. According to the conventional bulk-edge correspondence, these topological boundary modes are generally one dimension lower than the topological systems themselves. For instance, one-dimensional topological systems host zero-dimensional topological modes [9-11], while two-dimensional topological systems host one-dimensional topological edge modes [12-14]. These topological boundary modes exhibit robustness against various defects and perturbations, rendering them highly resilient. However, certain perturbations targeting their topological protection mechanisms can compromise their robustness. Strong symmetry-breaking or large-scale structural deformation are examples of such perturbations that can disrupt the robustness of the topological modes. For instance, edge modes of valley photonic crystals are resilient against perturbations that do not mix the two valleys [15, 16]. Hence, the sidewall roughness of valley photonic crystals can introduce backscattering and impact the robustness of topological edge modes [17].

Recent advancements have unveiled a novel class of topological phases known as higher-order topological phases, which deviate the traditional bulk-edge correspondence [18-22]. In contrast to conventional topological materials, the boundary modes of higher-order topological materials are more than one dimension lower than the bulk. For example, two-dimensional second-order topological photonic crystals (STPCs) can host zero-dimensional corner modes [23-35]. These corner modes are localized at the intersection of two or more lower-dimensional boundaries within the STPCs and hold significant promise for various applications, including on-chip cavities [36-38], advanced lasers [39, 40] and nonlinear optics [41, 42]. The corner modes are pinned at the zero frequency when the chiral symmetry is preserved. However, in realistic PC cavities, common perturbations tend to break the chiral symmetry, resulting in frequency shifts of the cavity modes. Moreover, the translation of rods or holes in PCs is a common fabrication error during the fabrication process and also induces frequency shifts of the corner modes. Recent theoretical proposals have shown that under translational deformation, gapless corner modes can be achieved by introducing synthetic dimensions [43]. The introduction of synthetic dimensions [44-46] in topological photonic systems enriches the exploration of topological phase of light in higher-dimensional space beyond three-dimensional real-space [47-50]. By considering the translational deformation as a synthetic parameter, topological boundary modes have been observed [51] and on-chip topological rainbow has been demonstrated [52]. However, the experimental observation of these gapless corner modes in synthetic translation dimensions is still absent, hindering the further applications in the design of microcavity with tunable working frequency.

In this work, we present the observation of gapless corner modes in PC slab under translational deformations. The PC slab consists of periodic ceramic square-rods placed on a metallic substrate. By introducing the translation ($\Delta x$, $\Delta y$) into the PC slab, we are able to tune the frequency of corner mode, thus demonstrating the existence of gapless corner mode in synthetic translation dimensions. To validate our findings, we employ the near-field scanning measurements to experimentally image the localized corner modes. We systematically explore and image for corner modes in PC slabs under different translational deformations, confirming the change of the frequency of corner mode. All experimental results are further supported by full-wave simulations, ensuring the reliability of our findings. Our research reveals that the corner modes in STPCs are actually the lower-dimensional manifestations of gapless corner modes. Moreover, our findings demonstrate the potential for the flexible frequency modulation of corner modes in synthetic translation dimensions.

**Gapless corner modes in synthetic translation dimensions**

To illustrate the presence of corner modes and their gapless dispersion in synthetic translation dimensions, we first consider the corner formed between a two-dimensional (2D) second-order topological photonic crystal (STPC) and an ordinary photonic crystal (OPC) [Fig. 1(a)]. The unit cell of the STPC consists of four square-rods with a side length of 3.5 mm. The distances between two neighboring rods in the in-plane directions are $d_x = d_y = 14.5$ mm. In contrast, the unit cell of the OPC consists of a square-rod with a side length of 7 mm. The lattice constant of both PCs is $a = 18$ mm and the relative permittivity of rod is 5.9. The topology of these two PCs is given by the 2D Zak phase **Z** = $(Z_x, Z_y)$ [24, 25]:

$$Z_j = \int dk_x dk_y Tr[\hat{A}_j(\boldsymbol{k})] \tag{1}$$

where $j = x$ or $y$ and $\hat{A}_j(\boldsymbol{k}) = i\langle u(\boldsymbol{k})| \partial_{k_j} |u(\boldsymbol{k})\rangle$ with $|u(\boldsymbol{k})\rangle$ being the periodic Bloch function. The 2D Zak phases of the STPCs and OPCs are $(\pi, \pi)$ and $(0, 0)$ respectively, enabling the localized corner modes when the STPCs and OPCs are placed together to form a corner. The eigenfrequencies of this corner configuration are calculated and presented in Fig. 1(b). The corner mode is denoted by red point, while the edge and bulk modes are denoted by blue and gray points, respectively. The inset of Fig. 1(b) shows the $|E_z|^2$ field of the corner mode, demonstrating strong field confinement at the corner. Despite the difference in unit cell between the STPC and OPC, the STPC's unit cell can be obtained by translating the dielectric rod within the OPC by $(\Delta x, \Delta y) = (a/2, a/2)$. This connection inspires the investigation of corner mode evolution during the dynamic process involving the translation of OPC. Hence, we replace the STPC with a translated OPC whose all square-rods are translated away from

the center of the unit cell by ($\Delta x$, $\Delta y$) [Fig. 1(c)]. For simplicity, we consider the case of $\Delta x = \Delta y$ and calculate the eigenfrequencies of corner modes with different $\Delta x$ values [denoted by red in Fig. 1(d)]. As $\Delta x$ varies from -9 to 9 mm, the corner modes bifurcate from the bulk band edge and traverse the entire photonic band gap. It is noteworthy that the corner modes highlighted by two green circles at $\Delta x = \Delta y = $ -9 and 9 mm are the same as the corner mode presented in Fig. 1(b). This indicates that the corner mode in the STPC is lower-dimensional manifestation of gapless corner modes under translational deformation. In fact, corner modes with tunable frequencies can also be obtained by changing $d_x$ and $d_y$ [Fig. S1]. However, the dispersion under variable $d_x$ and $d_y$ is limited to a small frequency range. The existence of gapless corner modes is determined by the topological pumping in synthetic translation dimensions [43], which is independent on specific boundary conditions. To further illustrate this point, we replace the OPC with a trivial perfect electric conductor (PEC) [Fig. 1(e)], and calculate its corner mode dispersion as a function of translation [Fig. 1(f)]. Remarkably, gapless corner modes that traverse the entire photonic band gap are still supported when $\Delta x = \Delta y$, and the complete frequency diagram of the corner modes in ($\Delta x$, $\Delta y$) space is shown in Fig. S2.

**Realization on photonic crystal slabs**

To observe gapless corner modes, we implement the above idea in the PC slab that utilizes both the in-plane periodicity and out-of-plane total reflection to confine light in three dimensions [Fig. 2(a)]. The unit cell of the PC slab consists of a dielectric square-rod with the relative permittivity of $\varepsilon_r = 9$ placed on a metallic substrate. The in-plane lattice constant is $a = 18$ mm, the side length and height of dielectric rods are $b = 7$ mm and $h = 13.5$ mm, respectively. Note that the bulk bands and eigen fields of this PC slab resemble those of the 2D PC presented in Fig. 1 [Fig. S3], indicating the potential existence of gapless corner modes in synthetic translation dimensions. To explore this possibility, we construct the "corner" by placing two metallic bars along the $x$ and $y$ directions next to the PC slab. We first calculate the eigenfrequencies for the sample with ($\Delta x$, $\Delta y$) = (-2 mm, -4 mm) [Fig. 2(b)]. Along the frequency axis, the bulk, edge and corner modes are denoted by gray, blue and red points, respectively. The frequency range of the band gap is outlined by a cyan rectangle. The $|E_z|^2$ fields at z = 0 mm for five representative bulk, edge, and corner modes are shown in Fig. 2(c). These fields show the frequency-dependent mode evolution. The bulk modes extend throughout the crystal, the edge modes propagate along $x$ (or $y$) direction while being confined along the other direction, and the in-gap corner mode with a frequency of 6.66 GHz is localized in both directions. To achieve gapless corner modes, we translate the PC slab with respect to two static metallic bars. For simplicity, we focus on the representative case of $\Delta x = \Delta y$. Figures 2(d)-2(f) show the eigenfrequencies and $|E_z|^2$ fields of corner modes for the samples with ($\Delta x$, $\Delta y$) = (-3 mm, -3 mm), ($\Delta x$, $\Delta y$) = (-4 mm, -4 mm) and ($\Delta x$, $\Delta y$) = (-5 mm, -5 mm). The frequency-dependent mode evolution for the samples with $\Delta x = \Delta y$ are shown in Fig. S4. In all samples, corner modes are found within the bulk band gap. In addition, the frequencies of corner modes increase from 6.67 GHz to 6.88 GHz to 7.07 GHz as $\Delta x$ decreases, showcasing the manifestation of gapless corner modes in PC slabs.

**Experimental observation**

The PC slab is free of the metallic cover and in favor for observation of eigen modes. To observe the corner modes, we demonstrate the experimental observation on the microwave near-field scanning platform [Fig. 3(a)]. The setup includes a source antenna and a probe antenna connected to a vector network analyzer (VNA) via a coaxial-cable [left panel]. The probe antenna, mounted on an automatic stepper motor, collects signals at $z = 15$ mm, which are sent back to the VNA and a computer for imaging the eigen modes. The sample consists of periodic ceramic rods and two metallic bars which are placed on the metallic substrate (top right panel). The corner of the sample is enlarged, and the source antenna is positioned at the center of four rods closest to the corner and is indicated by the blue circle (bottom right panel). We first consider the sample with $(\Delta x, \Delta y) = (-2$ mm$, -4$ mm$)$ to demonstrate the capability of imaging the eigen modes. The experimental $|E_z|^2$ fields of four representative eigen modes are shown in Fig. 3(b). At the frequency of 6.32 GHz, the bulk mode exhibits an extended field within the bulk PC slab. At 6.51 GHz, the $y$-edge mode shows a confined field localized around the $y$ edge, i.e., the boundary perpendicular to the $y$-axis. At 6.40 GHz, the mixed $x$&$y$ edge mode exhibits an extended field along both the $x$ and $y$ edges. This is induced by the field overlapping between the $x$ edge mode and the $y$ edge mode at the same frequency [see details in Fig. S5]. Importantly, at 6.71 GHz, a strongly confined field localized at the corner is observed, confirming the existence of corner mode. To validate these experimental results, we also perform the full 3D numerical simulation under the same excitation settings [Fig. 3(c)]. The simulated $|E_z|^2$ fields closely reproduce the experimental observations, providing further confirmation of the validity of the experimental results.

To further demonstrate the presence of gapless corner modes in synthetic translation dimensions, we perform measurements on samples with $\Delta x = \Delta y$. Since the source antenna is placed near the corner, there are field enhancements at frequencies where no eigen modes are expected. In order to accurately identify the resonantly excited corner modes, we define the response intensity $R = |E_z|^2_{\text{corner}}/|E_z|^2_{\text{source}}$, representing the ratio of the field intensity at the corner to that at the source. The positions of the corner and source are illustrated in the inset of Fig. 4(a). The response intensity reaches its maximum value when the corner mode is excited, and the peak of the response spectrum confirms the existence of the corner mode. Figure 4(a) shows the response intensity spectrum for the sample with (Δ*x*, Δ*y*) = (-3 mm, -3 mm), where the red line shows simulated result and the blue shows experimental result. The experimental and simulated frequencies of corner modes are 6.71 GHz and 6.70 GHz, respectively, demonstrating consistency between these two results. The measured and simulated $|E_z|^2$ fields of the excited corner modes for this sample are shown in Figs. 4(b) and 4(c), respectively, revealing a strongly localized field around the corner. Similarly, the response spectrum, as well as the measured and simulated $|E_z|^2$ fields, are presented for the sample with (Δ*x*, Δ*y*) = (-4 mm, -4 mm) in Figs. 4(d)-4(f), and for the sample with (Δ*x*, Δ*y*) = (-5 mm, -5 mm) in Figs. 4(g)-4(i). The frequencies of these corner modes are almost the same (with an error less than 0.8%), and the $|E_z|^2$ fields of these corner modes exhibit negligible differences, confirming the high consistency between the experimental and simulated results. By summarizing the results in Figs. 4(a), 4(d) and 4(g), we observe that the frequencies of corner modes increase from 6.71 GHz to 6.89 GHz and 7.17 GHz as (Δx, Δy) changes from (-3 mm, -3 mm) to (-4 mm, -4 mm) and (-5 mm, -5 mm), respectively. This provides clear evidence of the observation of gapless corner modes in synthetic translation dimensions.

**Conclusion**

Starting from the STPCs, we find that topological corner modes exhibit robust gapless dispersion in synthetic translation dimensions, and directly observe the gapless corner modes in PC slabs system. Specifically, utilizing the microwave near-field scanning platform, we map the mode evolution of the sample over a range of about 6-7 GHz from bulk mode to topological corner mode within the band gap. We verify the gapless characteristic of corner modes by changing the global translation of rods and measuring response intensities. All measured results are in good agreement with the simulated results. The demonstrated gapless corner modes under synthetic translational deformations is universal and can be applied to other lattice structures. Our study demonstrates a feasible strategy for accessing and tuning topological corner modes at microwave frequencies. Although our implementation involved a metal-containing microwave system, it can be extended to nanophotonic systems working at optical frequencies. This work shows the potential application of topological corner modes in nonlinear optics and quantum optics.

**Acknowledgements**

This work was supported by State Key Research Development Program of China (2022YFA1404304), National Natural Science Foundation of China (12074443, 62035016), Guangdong Basic and Applied Basic Research Foundation (2019B151502036, 2023B1515040023), Fundamental Research Funds for the Central Universities, Sun Yat-sen University (Grant No. 23lgbj021).

**Competing financial interests**

The authors declare no competing financial interests.

**Code, Data, and Materials Availability**

The data that support the findings of this study are available from the corresponding authors upon reasonable request.

**Biographies**

**Wen-Jin Zhang** is currently pursuing his Master of Philosophy at the School of Physics, Sun Yat-sen University. His research interests include topological photonics based on photonic crystals.

**Hao-Chang Mo** is currently pursuing his Master of Philosophy at the School of Physics, Sun Yat-sen University. His research interests include topological photonics based on photonic crystals.

**Wen-Jie Chen** received his PhD from School of Physics and Engineering of Sun Yat-sen University,

China, in 2014. He is currently a professor at the School of Physics of Sun Yat-sen University. His current research interests include topological photonics based on photonic crystals and metamaterials.

**Xiao-Dong Chen** received his PhD from School of Physics of Sun Yat-sen University, China, in 2016. He is currently an associate professor at the School of Physics of Sun Yat-sen University. His current research interests include topological photonics and nanophotonics based on photonic crystals and metamaterials.

**Jian-Wen Dong** received his PhD from School of Physics and Engineering of Sun Yat-sen University, China, in 2007. He is currently a professor at the School of Physics of Sun Yat-sen University. His current research interests include topological photonics, nanophotonics, metasurfaces, 3D holography.

# Figures and captions

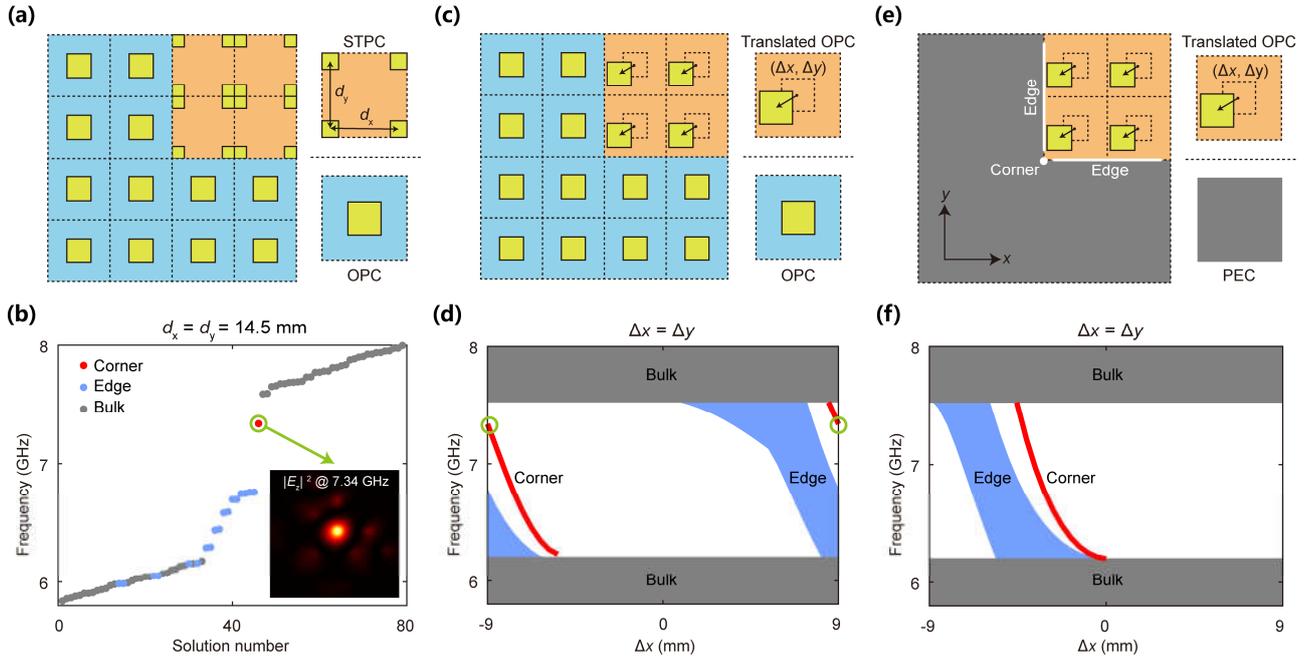

**Figure 1 | Corner mode and its gapless dispersion in synthetic translation dimensions. (a)** Schematic illustration of the corner between STPC and OPC. Parameters: lattice constant $a$ = 18 mm, the side length of square-rod of OPC (blue region) is 7 mm, while that of STPC (orange region) is 3.5 mm, in-plane distances between two neighboring rods in STPC are $d_x = d_y$ = 14.5 mm, relative permittivity of rods $\varepsilon_r$ = 5.9. **(b)** Calculated eigenfrequencies of configuration in (a). The corner, edge and bulk modes are denoted by red, blue and gray points, respectively. Inset: The simulated $|E_z|^2$ field of the corner mode (outlined by a green circle). **(c)** Schematic illustration of the corner between the translated OPC and the untranslated OPC. All square-rods of the translated OPC are translated away from the center of the unit cell by ($\Delta x$, $\Delta y$). The STPC corresponds to the translated OPC with $\Delta x = \Delta y = a/2$ = 9 mm. **(d)** The gapless dispersion of corner modes of configuration in (c) when $\Delta x = \Delta y$. Corner modes outlined by two green circles at $\Delta x = \Delta y$ = -9 and 9 mm are the same as the corner mode in (b). **(e)** Schematic illustration of the corner between the translated OPC and PEC. **(f)** The gapless dispersion of corner modes of configuration in (e) when $\Delta x = \Delta y$.

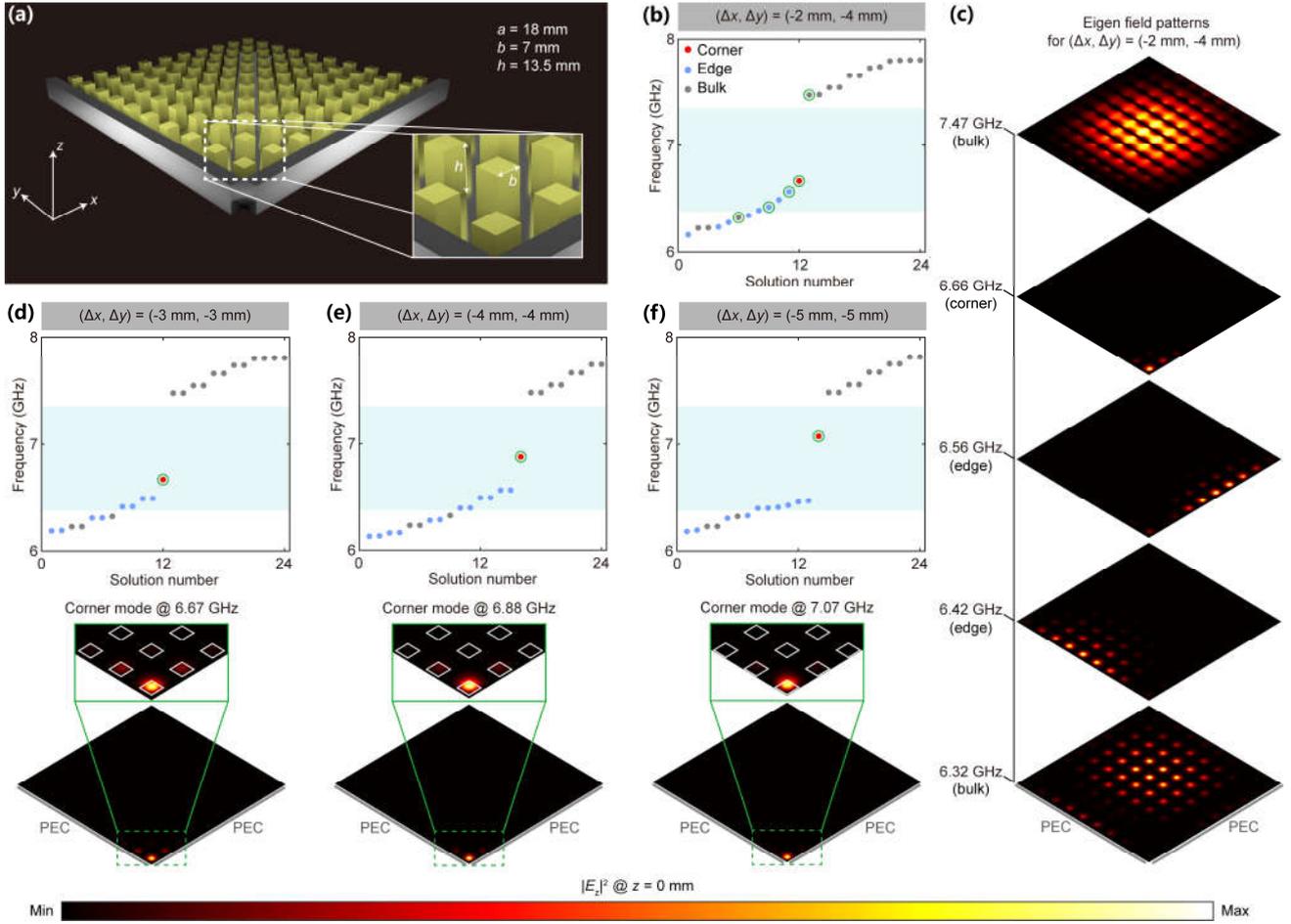

**Figure 2 | Realization of gapless corner modes based on PC slab. (a)** Schematic illustration of the corner between the PC slab and two metallic bars. Both the PC slab and metallic bars are placed on a metallic substrate. Parameters: lattice constant $a$ = 18 mm, in-plane side length of dielectric rods $b$ = 7 mm, height of dielectric rods $h$ = 13.5 mm and relative permittivity of dielectric rods $\varepsilon_r$ = 9. **(b)** Calculated eigenfrequencies for the sample with $(\Delta x, \Delta y)$ = (-2 mm, -4 mm). The cyan region represents the bulk band gap of PC slab. The eigen modes whose $|E_z|^2$ fields will be shown in (c) are outlined by circles. **(c)** $|E_z|^2$ fields at z = 0 mm for bulk, edge, and corner modes in (b). **(d-f)** Calculated eigenfrequencies and $|E_z|^2$ fields of the corner modes for the sample with $(\Delta x, \Delta y)$ = (-3 mm, -3 mm), $(\Delta x, \Delta y)$ = (-4 mm, -4 mm) and $(\Delta x, \Delta y)$ = (-5 mm, -5 mm).

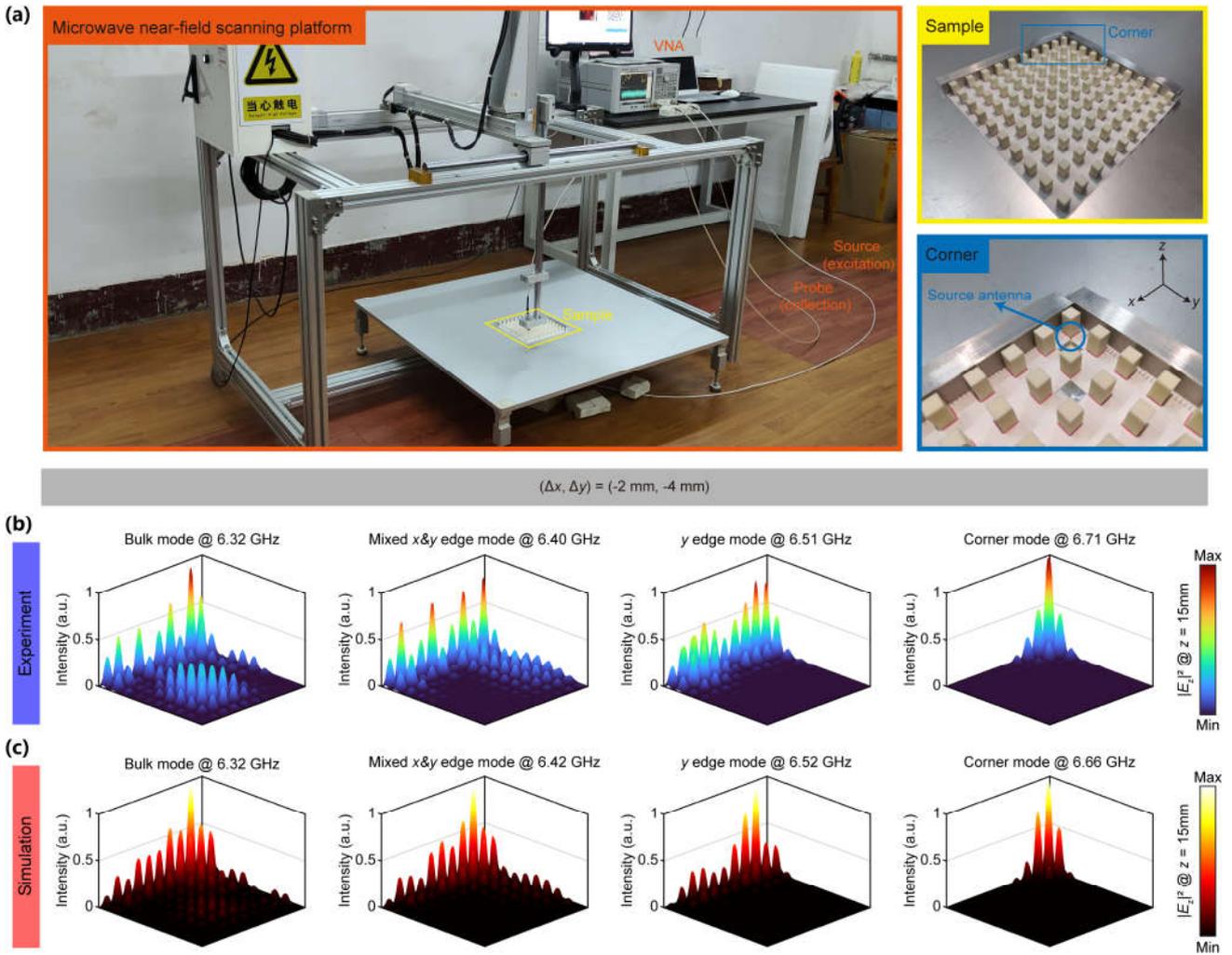

**Figure 3 | Observation of eigen modes of the corner between the PC slab and two metallic bars.**
**(a)** Photographs of microwave near-field scanning platform and sample used in the measurement. Left panel: the source antenna connected to VNA is used to excite the input source, and the probe antenna above the sample is used to measure the electric field and mounted on a motorized translation stage. Top right panel: periodic ceramic rods and two metallic bars are placed on a metallic substrate, constructing the sample supporting in-gap corner modes. Bottom right panel: the corner of sample is enlarged and a blue circle marks the source antenna. **(b)** Experimental $|E_z|^2$ fields at $z = 15$ mm of four eigen modes for the sample with $(\Delta x, \Delta y)$ = (-2 mm, -4 mm), representing the bulk mode, the edge mode and the corner mode. **(c)** Simulated $|E_z|^2$ fields of four representative eigen modes, which are consistent with the experimental results.

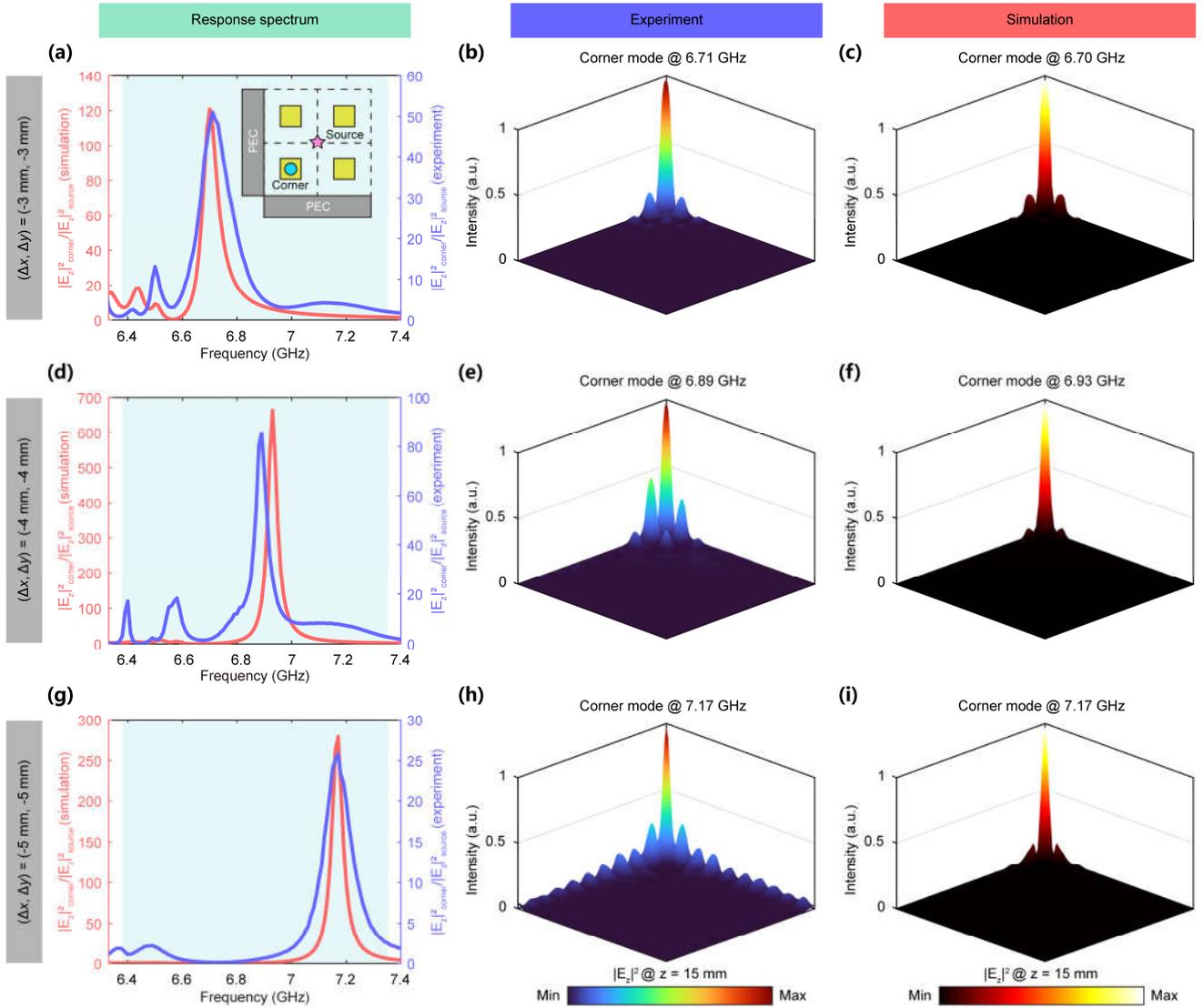

**Figure 4 | Observation of gapless corner modes within the translation dimensions. (a)** Response spectrum for the sample with ($\Delta x$, $\Delta y$) = (-3 mm, -3 mm). The red line is the simulated result and the blue is the experimental result. The cyan region represents band gap. The response intensity is defined as $|E_z|^2_{corner}/|E_z|^2_{source}$, whose peak indicates the excitation of corner mode. For the sample with ($\Delta x$, $\Delta y$) = (-3 mm, -3 mm), the central frequencies of corner modes are 6.70 GHz (simulation) and 6.71 GHz (experiment). Inset: illustration of two positions at which electric field is used to determine the response intensity. **(b, c)** Measured and simulated $|E_z|^2$ fields of the excited corner mode for the sample with ($\Delta x$, $\Delta y$) = (-3 mm, -3 mm). **(d-f)** and **(g-i)** are similar to **(a-c)**, while (d-f) correspond to the sample with ($\Delta x$, $\Delta y$) = (-4 mm, -4 mm) and (g-i) correspond to the sample with ($\Delta x$, $\Delta y$) = (-5 mm, -5 mm).